\documentclass[journal,transmag,comsoc,final]{IEEEtran}

\usepackage[T1]{fontenc}
\usepackage{cite}
\usepackage{url}
\usepackage{amsmath}
\usepackage{amssymb}
\usepackage[cmintegrals]{newtxmath}
\usepackage{subcaption}
\usepackage{dblfloatfix}
\usepackage{graphicx}
\usepackage{epstopdf}
\usepackage{xcolor}
\usepackage{mathtools}
\usepackage[detect-weight=true, binary-units=true]{siunitx}
\usepackage[inline]{enumitem}
\usepackage{eurosym}
\usepackage{adjustbox}

\newcommand{\revise}[1]{\textcolor{\highlightcolor}{#1}}
\renewcommand{\revise}[1]{#1}

\begin{document}
%
\title{D2D-Based Grouped Random Access to Mitigate\\ Mobile Access Congestion in 5G Sensor Networks}
%
%
%

\author{Bin~Han,~\IEEEmembership{Member,~IEEE,}
        Vincenzo~Sciancalepore,~\IEEEmembership{Member,~IEEE,}
        Oliver~Holland,~\IEEEmembership{Member,~IEEE,}\\
        Mischa~Dohler,~\IEEEmembership{Fellow,~IEEE,}
        and~Hans~D.~Schotten,~\IEEEmembership{Member,~IEEE}
\thanks{\textit{The work of V. Sciancalepore was supported by the European Union H-2020 Project 5G-TRANSFORMER under grant agreement no. 761536.}}
\thanks{\textit{B. Han and H. D. Schotten are with {Technische Universit\"at Kaiserslautern}.}}
\thanks{\textit{\{binhan,schotten\}@eit.uni-kl.de}}%
\thanks{\textit{V. Sciancalepore is with NEC Laboratories Europe.}}
\thanks{\textit{vincenzo.sciancalepore@neclab.eu}}%
\thanks{\textit{O. Holland and M. Dohler are with King's College London.}}
\thanks{\textit{\{oliver.holland,mischa.dohler\}@kcl.ac.uk}}%
}

\newcommand\blfootnote[1]{%
	\begingroup
	\renewcommand\thefootnote{}\footnote{#1}%
	\addtocounter{footnote}{-1}%
	\endgroup
}

%
%

\ifCLASSOPTIONpeerreview
\else
\fi
%





\maketitle
\setlength{\textfloatsep}{5pt}

\begin{abstract}
The Fifth Generation (5G) wireless service of sensor networks involves significant challenges when dealing with the coordination of ever-increasing number of devices accessing shared resources. \revise{This has drawn major interest from the research community as many existing works focus on the radio access network congestion control to efficiently manage resources in the context of device-to-device (D2D) interaction in huge sensor networks.}
In this context, this paper pioneers a study on the impact of D2D link reliability in group-assisted random access protocols, by shedding the light on beneficial performance and potential limitations of approaches of this kind against tunable parameters such as group size, number of sensors and reliability of D2D links. Additionally, we leverage on the association with a Geolocation Database (GDB) capability to assist the grouping decisions by drawing parallels with recent regulatory-driven initiatives around GDBs and arguing benefits of the suggested proposal. Finally, the proposed method is approved to significantly reduce the delay over random access channels, by means of an exhaustive simulation campaign.
\end{abstract}


%
\IEEEpeerreviewmaketitle

\thispagestyle{empty}	
\section{Introduction}\label{sec:intro}

\begin{figure}[!htb]
	\centering
		\fbox{\label{key}\includegraphics[width=.47\textwidth]{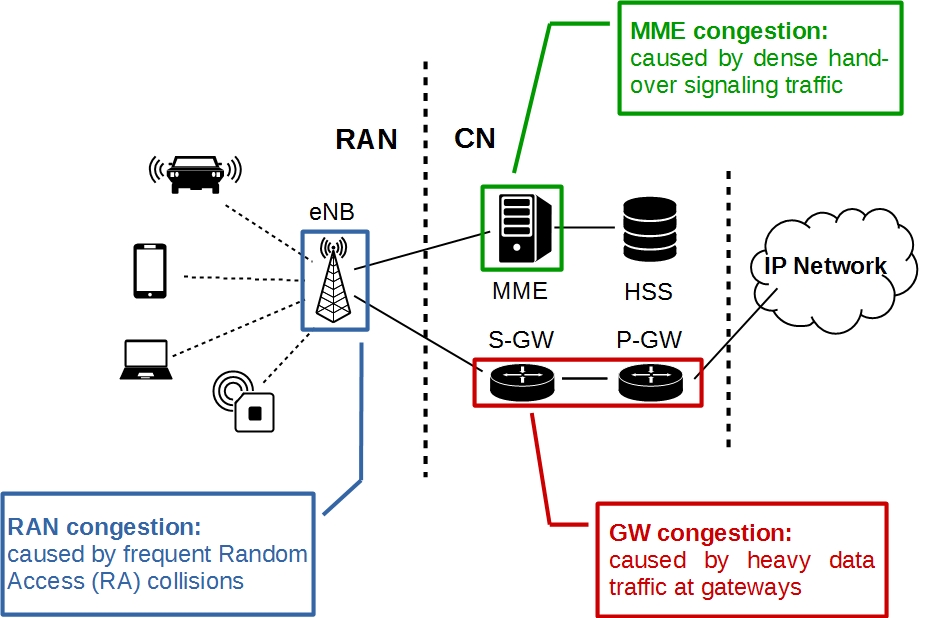}}
	\caption{Congestions on different domains in legacy networks.}
	\label{fig:congestion_types}
\end{figure}

\IEEEPARstart{T}{he} plethora of small, smart sensors becoming available and the range of interesting applications using them are changing our day-to-day life, and calling for novel technological solutions able to shake up the wireless communication landscape~\cite{soldani2015horizon}.
{When such a huge number of devices need inter-connectivity and low access delay, additional technical challenges must be addressed to avoid traffic congestion and service degradation. In particular, network congestion might occur at different levels due to several reasons. The current Long-Term-Evolution (LTE) and LTE-Advanced (LTE-A) deployments might exhibit vulnerability to data congestion at $i$) the Radio Access Network (RAN), $ii$) the Mobility Management Entity (MME), and, $iii$) the core network GateWays (GWs), including the Serving Gateway (S-GW) and Packet Data Network Gateway (P-GW).} This is illustrated in Fig.~\ref{fig:congestion_types}. 

RAN congestions are due to uplink random access (RA) collisions emphasizing the huge sensitivity to the device density. MME congestion is caused by handover signaling overloads, e.g. due to nomadic users, hence showing sensitiveness to high user mobility. GW congestion has roots in a shortage of gateway capacity and heavy data traffic, showing a strict dependency on the overall user data-rate.

\blfootnote{This is a preprint, the full paper has been accepted by \emph{IEEE Communciations Magazine}, \copyright 2019 IEEE. Personal use of this material is permitted. However, permission to use this material for any other purposes must be obtained from the IEEE by sending a request to pubs-permissions@ieee.org.}

Compared to human-type communication (HTC), new-generation sensor networks---as a typical application of massive machine-type communications (mMTC)---exhibit different behaviors. First, the density of sensors in the deployment area can be substantially higher than handheld HTC devices (HTCDs). Second, most sensors are characterised by quasi-static mobility, compared with high mobility in HTCDs. This may allow for predictable information and accurate scheduling policies as explained hereafter. Additionally, human users usually exchange among each others (or directly to a cloud entity) large amounts of data in each communication session. For example, a voice call usually lasts several minutes, generating megabytes of data through common Voice-over-IP codecs
. The size of an email, as another example, typically ranges from dozens of \SI{}{\kilo\byte}s to several \SI{}{\mega\byte}s, depending on the attachment. In contrast, sensor messages often have a small payload size of less than \SI{1}{\kilo\byte}\cite{3gpp2014geran}. Besides, a special behavior can be observed in sensor networks: sensors are usually synchronized to the same schedule of data transmission thereby generating periodical bursts of random access (RA) requests in the RAN, which rarely happens in the context of HTCDs. Considering these peculiarities, we can argue that \emph{sensor networks are severely impaired by RAN congestion, rather than MME or core-gateway congestions}.

Great effort has been invested in reducing the RA collisions in sensor networks, as detailed in~\cite{hasan2013random,ali2017lte}. During recent years, {encouraged by the popularity of deploying device-to-device (D2D) communications and unlicensed bands in cellular systems \cite{lien20163gpp,wu2016device
}, a variety of new RAN congestion control approaches based on D2D communications and device grouping have attracted much interest from academic and industrial players}, due to their advantages in energy consumption and access delay~\cite{sciancalepore2016offloading}. However, these methods generally assume D2D links as reliable, which can {rarely} be argued in practice. Besides, they generally only focus on the radio resource management of the physical layer, and lack discussion about the protocol design on the medium access control (MAC) and higher layers. In this work, we rely on the unreliability of D2D links, and thus enhance the \revise{existing approaches in this category introducing both a protocol extension and an architectural solution to efficiently handle access congestion issues in 5\textsuperscript{th} Generation (5G) sensor networks. Main contributions can be summarized as the following: $i$) we propose a new MAC frame design, $ii$) we detail a novel hierarchical geolocation database (GDB) signaling approach and $iii$) we blend both solutions together to enable a satisfactory initial clustering and an online group updating, towards a more reliable and efficient RA solution.}


\section{mMTC RAN Congestion Control: State-of-the-Art}\label{sec:sota}


A number of approaches have been explored and designed to properly control RAN congestions in mobile networks, as they might severely impair the overall network efficiency. 
Surveys with a deep comparative study on these approaches have been conducted in~\cite{hasan2013random,ali2017lte}. \revise{We have analyzed and briefly summarized in the following the main findings that allow us to highlight the differences against our novel approach.}
\begin{itemize}
	\item \textbf{Access Class Barring (ACB)}: ACB check is performed to accept/reject every sensor device. Upon rejection, the device waits for a given timeout before re-attempting. Admission probability and timeout length depend on the access class of device. However, advanced methods have been suggested to improve such a solution.
%
	\item \textbf{Prioritized RA}: Applications and Random Access Channels (RACHs) are divided into different classes to optimize the RACH resource allocation w.r.t. required Quality-of-Service (QoS).
	\item \textbf{MTC-Specified Backoff}: A backoff mechanism is implemented to prevent a user equipment (UE) from colliding with other contending UEs.
	\item \textbf{RACH Resource Separation}: RACH resources are optimally assigned beforehand to HTC/MTC applications.
	\item \textbf{Dynamic RACH Allocation}: A RACH resource allocation is dynamically performed at base station (BS) based on instantaneous congestion levels.
	\item \textbf{Pull-Based RA}: Base stations proactively grant sensor devices to access without waiting for RA requests. This solution comprises different schemes: $i$) Pull-Based Individual Paging, $ii$) Pull-Based Group Paging and $iii$) Pull-Based Group Access.
	\item \textbf{Self-Optimization Overload Control RA}: A combination of RACH resource separation, dynamic RACH allocation and ACB is realized. 
	\item \textbf{Code-Expanded RA}: A preamble set is issued instead of a single preamble for any RA request. 
	\item \textbf{Spatial-Grouping}: Spatial diversity is introduced to reduce collisions and increase the RA preamble reuse.
	\item \textbf{Guaranteed RA}: RA load is estimated during the RA procedure to boost the control scheme optimization.
	\item \textbf{Non-Aloha-Based RA}: Multiple-access is combined with resource allocation based on analog fountain code.
\end{itemize}

However, above-mentioned solutions present limitations when the number of sensors dramatically increases.
A plausibile solution brings into play the clustering concept, where sensor devices are promptly grouped while exploiting D2D transmissions for intra-cluster data exchange\,\footnote{D2D transmissions are usually performed by means of short-range communications, such as Bluetooth, WiFi-Direct and ZigBee. Recently, mm-wave radio technologies are being explored to realize interference-less communications, such as WiGig D2D. However, they still present technological limitations and are not considered in this work.}. We call such solutions collectively as \textit{D2D-based grouped RA} (D2D-GRA). 
%

{Differing from other methods, the concept of D2D-GRA was not originally motivated to control RAN congestion, but to reduce power consumption~\cite{ji2017applying}.}
This kind of solution drives the system to energy-efficient states, especially attractive for battery-life-critical mMTC applications, such as new-generation sensor networks. Interestingly, they are proven efficient in RA collision reduction as demonstrated in~\cite{chatzikokolakis2015way}. When combined with device classification, this class of techniques also benefits from guaranteeing a very low access delay for periodical devices of certain classes, which is valuable in duty-cycle-critical use cases such as time-constrained wireless sensor networks. \revise{However, a well-known drawback is the huge amount of cluster management operations for devices with high mobility, though it is not a critical aspect in common sensor networks where the devices (sensors) are quasi-static or with very-low mobility.}

{Despite of all these advantages, yet a fundamental question shall be answered before entitling D2D-GRA as the most promising solution for RAN congestion control in sensor networks: \emph{Is the D2D link reliability an essential feature while enabling the intra-group data aggregation and distribution}?}
All existing methods are developed based on the same assumption, i.e., D2D links are fully available and reliable. This might bias the performance evaluation of real deployments as such an assumption is not always feasible in practice. \revise{Our proposal pioneers a novel protocol that is able to efficiently deal with D2D exceptions}.

\section{D2D Link Reliability in Grouped RA}\label{sec:d2d}
\begin{figure}[!htb]
	\centering
	\fbox{\includegraphics[width=.47\textwidth]{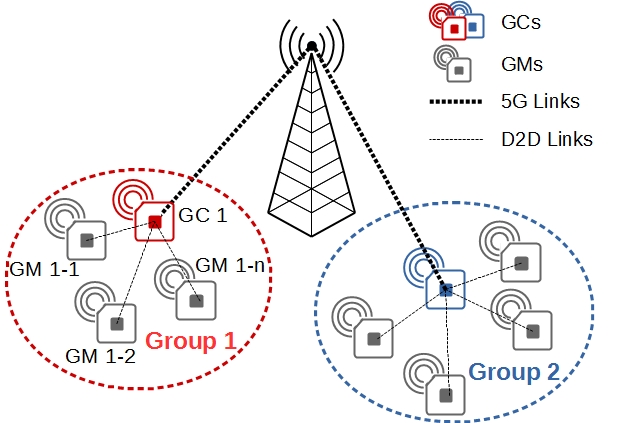}}
	\caption{RAN topology in D2D-GRA.}
	\label{fig:ran_topology_grouped_ra}
\end{figure}

The principle of D2D-GRA is illustrated in Fig.~\ref{fig:ran_topology_grouped_ra}. Sensors are clustered into groups, generally according to their spatial locations, however context information such as device class can also be accounted. Each group has a group coordinator (GC), which relays data for the rest group members (GMs). To accomplish such relaying operations, a D2D link must be established between the GC and every GM, such that the GC can aggregate uplink data from the GMs, and distribute downlink data to them. \revise{For the GMs without direct physical access to the BS, the end-to-end connection management shall be executed at the BS with assistance of their dedicated GCs, and a transport connection shall be established as introduced in \cite{soldani2008wireless}.} Thus, in each group, only the GC executes the RA procedure to visit the BS so that the RA request density---which is inversely proportional to the average group size---is reduced. While an extra overhead may be generated by the essential intra-group signaling messages, its impact on the macro cell RAN can be minimized or even removed by exploiting unlicensed spectrum technologies, namely D2D outbound communication, by means of e.g. Bluetooth or WiFi Direct. 

To further achieve an optimal energy efficiency, the processes of clustering and GC selection are usually carried out at the BS level, where cell-wide Channel Status Information (CSI) is available to support centralized scheduling.
Besides, as the energy consumption of GC might be unbalanced  due to additional coordination activities, the GC selection process might rely on the historical GC assignments so as to provide fairness guarantees.
An example of such history-based GC selection scheme has been reported in~\cite{ji2017applying}, where a D2D-based clustered uplink transmission solution is proposed for Internet-of-Things deployments.

However, as indicated earlier, the intra-group D2D links are not guaranteed to be always available or reliable. Due to interference and channel losses, some GMs may fail to connect to their GCs as assigned by the BS. Moreover, even established D2D links might vanish due to mobility and channel fading. When such an exception occurs, the involved GM becomes unable to communicate with the BS.  \revise{Although this issue can be resolved at some degree by deploying techniques such as automatic repeat query (ARQ), a significant compromise in delay or blocklength will occur as the cost. In case of stringent time requirements, the impact of D2D link exceptions becomes critical thereby challenging the overall D2D-GRA design.}

\begin{figure}[!htb]
	\centering
	\fbox{\includegraphics[width=.47\textwidth]{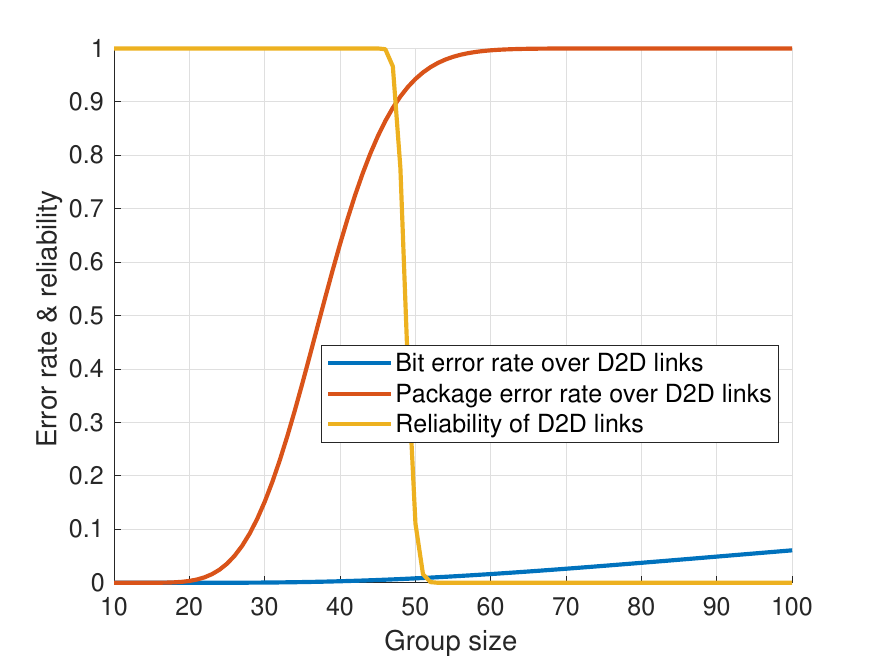}}
	\caption{{D2D performance is sensitive to sensor group size.
		}}
	\label{fig:d2d_reliability}
\end{figure}

The occurrence of exceptions can increase along with the group size, mainly due to a two-fold reason. The former is that, under the same spacial distribution of sensors, the geographical diameter of the group increases with the group size, leading to a larger average distance between the GC and its GMs, and hence a lower average signal-to-noise ratio (SNR). 
The latter, given the amount of radio resources for the GC to aggregate/distribute data to/from its GMs, for each link both the average channel capacity and the opportunities for retransmission will sink as the group grows. A simplified performance analysis of Bluetooth Low Energy (BLE) clusters present a logistic increase of packet error rate (PER) and a logistic decrease of D2D link reliability w.r.t. the group size, as shown in Fig. \ref{fig:d2d_reliability}. 

We can assert that the D2D connections are not always reliable, especially for large groups. Furthermore, there is a critical mass of the group size that dramatically reduces the D2D link reliability to a poor level. On the one hand, as suggested by Fig.~\ref{fig:d2d_reliability}, if D2D devices are clustered into groups of limited sizes, an ultra-high reliability is still achievable. On the other hand, having unnecessarily small groups will reduce the efficiency of GRA. Therefore, the optimal size can be identified as a trade-off that depends on various factors including the context information of sensors, especially the D2D CSI, which must be obtained through device measurements. For the initial clustering, due to the lack of D2D CSI, the optimum is usually hard to achieve. To demonstrate the impact of CSI estimation error on initial clustering performance, we conducted a simulation campaign under the reference scenario of urban coverage considering massive connection~\cite{3gpp2015study}, with over $8000$ synchronized sensors organized in $50$ clusters connected through BLE for D2D-GRA.

\begin{figure}[!htb]
	\centering
	\fbox{\includegraphics[width=.47\textwidth]{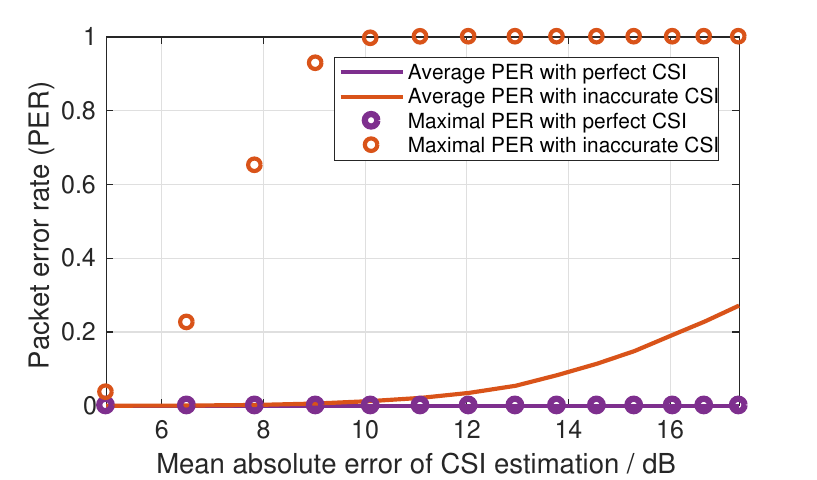}}
	\caption{The D2D links highly rely on the channel estimation.}
	\label{fig:PER_to_CSI_error}
\end{figure}

The results in Fig.~\ref{fig:PER_to_CSI_error} indicate that poor CSI estimations lead to a significant degrade of D2D links reliability. More specifically, the worst-case performance is much more sensitive than the average reliability to such errors. When the mean absolute estimation error exceeds \SI{6}{\dB}, at least one sensor suffers from a PER over $20\%$, although the average PER is still negligible. When further raising this error to \SI{10}{\dB}, the average PER increases to around $1\%$, while the worst clustered device fails almost every transmission. Although a non-optimal clustering can be recursively improved through global or local group updates, e.g. the methods proposed in \cite{han2017group}, the convergence process can be slow and showing a significant amount of link exceptions. Besides, an inappropriate initialization may lead to early convergence at sub-optimum instead of the global optimum. Moreover, once a link exception occurs, the BS shall be informed to update its clusters list, and the sensors in D2D link exception should reattempt to access the BS. One straightforward solution is to allow the sensor to send an extra RA request directly to the BS as an un-clustered device. However, when under a high D2D link exception rate, a huge amount of extra RA requests can be generated, which may even eradicate the gain brought about by D2D-GRA, as demonstrated in~\cite{han2017group}.

Therefore, two challenges shall be handled to better exploit the gain of D2D-GRA in 5G sensor networks: \textit{\begin{enumerate*}
		\item Designing a more efficient RAN resource assignment approach to detect and handle D2D link exceptions;
		\item Leveraging on external data sources, such as the geolocation databases (GDBs), to get a-priori knowledge about the channels thereby assisting the initial clustering.
\end{enumerate*}}


\section{Enhanced Grouped RA Protocol}\label{sec:protocol}
To efficiently setup and update sensor groups, three kinds of operations are essential in GRA\footnote{We refer the readers to~\cite{han2017group} for a detailed dissertation on these processes.}:
\begin{itemize}
	\item\textbf{Global group update}, which reclusters all sensors in the cell and selects a GC for every group. This operation is executed for initial clustering, and occasionally repeated to guarantee a low collision rate.
	\item\textbf{Group joining}, which allows a new device to join an existing sensor group, and eventually reselects the coordinator of updated group, if necessary. This operation can be triggered by handover or device attachment.
	\item\textbf{Group leaving}, which removes a device from its group, and eventually reselects the coordinator of involved group, if the leaving device is the current GC. This operation can be triggered by $i$) handover or device detachment, $ii$) D2D link exception of a GM and, $iii$) macro cell link collapse of the GC.
\end{itemize}

\begin{figure*}[!htb]
	\centering
	\fbox{\includegraphics[width=\textwidth]{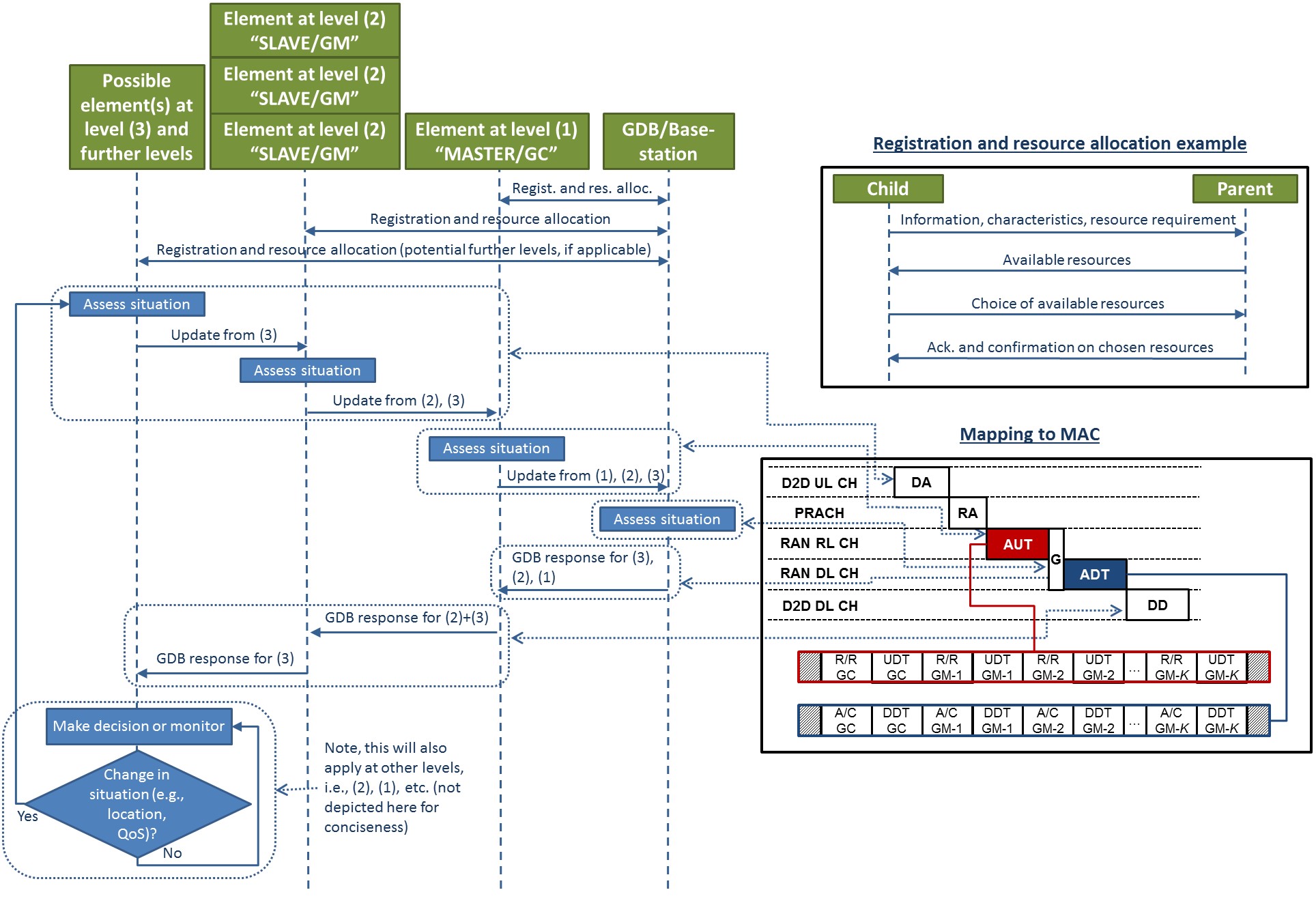}}
	\caption{\revise{The MAC frame for our D2D-GRA protocol, which is described in Sec.~\ref{sec:protocol}, is seamlessly integrated with the GDB signaling approach explained in Sec.~\ref{sec:gdb}.} The terms R/R, UDT, A/C and DDT denote request/report, uplink data, acknowledgment/command and downlink data, respectively.}
	\label{fig:gdb}
\end{figure*}

Clearly, besides the user data, extra signaling traffic is generated through this procedure, including requests for group leaving and reports of D2D link status in uplink, as well as request acknowledgments and commands for group updating/joining in downlink. The performance of D2D-GRA seriously decreases, if all these messages, especially the uplink ones, are transmitted in extra sessions. To avoid this, we design an enhanced transmission frame structure that integrates the signaling overhead with the user data payload, as illustrated in Fig.~\ref{fig:gdb} (bottom-right).
In this method, each complete transmission cycle of a sensor group consists of six time slots:
\begin{itemize}
	\item \textbf{Data Aggregation (DA)}, wherein the GC aggregates uplink user data from its GMs through the D2D uplink channels. Upon the applied D2D technology, D2D channel measurements are also executed during this phase.
	\item \textbf{Random Access (RA)}, wherein the GC attempts to access the BS via the Physical Random Access Channel.
	\item \textbf{Aggregated Uplink Transmission (AUT)}, wherein the aggregated uplink user data are packaged together with user requests and reports into one packet, and sent by the GC to the BS over the RAN uplink channel. The uplink packet format is depicted in Fig.~\ref{fig:gdb}. Depending on the system complexity requirement, the length of every packet segment can be either fixed or floating.
	\item \textbf{Guard (G)}, which provides a guarding interval. It also reserves processing time for the group management algorithm running at the BS. Pilot symbols for channel measurement of the GC are also transmitted therein.
	\item \textbf{Aggregated Downlink Transmission (ADT)}, wherein the aggregated downlink user data are packaged together with request acknowledgments and controlling commands into one packet, and sent by the BS to the GC over the RAN downlink channel. The downlink packets have a very similar format to the uplink packet, see Fig.~\ref{fig:gdb}.
	\item \textbf{Data Distribution (DD)}, wherein the GC distributes the received downlink user data to its GMs through the D2D downlink channels.
\end{itemize}

Signaling overhead for the D2D-GRA procedure is then carried along the user plane, without incurring in extra RA requests. However, the integrated overhead extends the length of aggregated data packets and increases the overall data traffic in RAN. Nevertheless, as discussed in Section~\ref{sec:intro}, sensor network applications generally generate very limited data traffic, and the risk of RAN congestion is much more critical than GW congestions. Therefore, a reasonable increase of data traffic is an acceptable trade-off for the collision reduction.


\section{Geolocation-Database-Assisted Clustering}\label{sec:gdb}
A final important aspect of our proposed solution is the use of geolocation and advanced context information (e.g., propagation mapping). This is to manage the grouping of sensors through centralized management and decision making. The use of a Geolocation Database (GDB) in this context, along with appropriate messaging built into the proposed protocol, serves a number of purposes and benefits. Particularly, it facilitates:

\begin{enumerate}
\item The optimal and efficient sensor clustering based on the D2D channel conditions, taking into account locations, propagation, the requirements in terms of communication characteristics and data rate, among others.  It also achieves a system-level viewpoint considering, e.g., interactions among the transmissions of different groups.
\item Collecting information from sensors to enhance the GDB in serving the above purpose, as well as others.
\end{enumerate}

Indeed, although an initial locus in the use of geolocation as a wireless communication facilitator has been location-based spectrum sharing and regulatory considerations, there are numerous other potential benefits and justifications to use this concept in 5G contexts. This is for reasons such as:
\begin{enumerate}
\item 5G technologies \emph{will} require spectrum sharing to achieve sufficient spectrum availability in some scenarios, e.g., at lower frequencies for coverage/reliability and signaling/control purposes.
Such spectrum sharing will often be between very different services and spectrum owners, such that the sharing must usually be approved by the regulator---typically through automated regulatory-driven or certified GDBs. A number of regulatory initiatives/trials are being undertaken or have been completed showing such GDB concepts in action, including TV white space (TVWS) \cite{hol2016}, Licensed-Shared Access \cite{mat2014}, and the Citizens Broadband Radio Service \cite{CBRS2015}---for which an LTE band is already designated.

\item One key aspect of 5G is heterogeneity; moreover, common management among the heterogeneous systems/elements in 5G will likely not preexist. A GDB can assist in managing connectivity, QoS, and other aspects in heterogeneous scenarios. {Furthermore, spectrum sharing/management at higher levels can also be integrated through combination with of regulatory-run or approved GDBs.} This is inherently possible because the GDB(s) will (at least partially) operate at higher (e.g., regulatory---transcending spectrum services and owners) levels for above-mentioned reasons.
Such databases (e.g, TVWS databases) under regulatory approval often incorporate aspects that can be reused for other management purposes. These include advanced propagation and context (e.g., transmitter and receiver locations and characteristics) knowledge, facilitating the management of QoS and associated allocation, connectivity, and other aspects in heterogeneous networking scenarios.

\item In many 5G and beyond contexts, it is necessary to consider the precise locations of equipment in resource management. This applies not only to spectrum resources, but also to computational resources achieving network functions through virtualization. One example here is the context of latency reduction, where careful geographical/location placing/instantiation of virtualized equipment is required by network management in order to minimize propagation delay, achieving as direct as possible propagation path between the communication endpoints.
\end{enumerate}

Fig.~\ref{fig:gdb} provides a GDB signaling example derived from a regulatory spectrum sharing GDB such as in TVWS or CBRS. Such a signaling begins with a registration and initial resource allocation procedure \revise{(top-right of Fig.~\ref{fig:gdb})}, which might include the transfer of technical characteristics, location and resource requirements. This GDB signaling is conceptually transferable and integrable to wireless RA procedure, in which the ``parent'' entities are the BSs and their dedicated GDBs whereas the ``child'' nodes are sensors that communicate directly with the BS for resources. More specifically, integrated with the grouped RA concept, the sensors will be eventually designated as GCs and GMs, or in case of necessity also subordinates in further level(s), \revise{which is the aforementioned initial clustering. In addition to enabling D2D-GRA, this also provisions for the option---with respect to the GDB system---of a GC (MASTER) acting on behalf of its GMs, if, e.g., they are not able to establish a connection directly to the GDB~\cite{hol2016}.}

Based on this initial exchange, the GDB will already have detailed information on sensors' locations. GDBs, for a range of purposes (e.g., TVWS, CBRS, etc.) also typically have advanced information on propagation/loss and other radio characteristics on a per-location basis, able to ascertain far better detail on the channel sets \emph{between the sensors} than a BS could otherwise do---in this latter case, only having information on the channels between the BS and each of the sensors. \revise{Shared with such information by the GDBs, BSs become capable to efficiently setup the initial sensor clusters.} Thereafter, the GC and GMs respectively assume the parallels of MASTER and SLAVE sensors using the TVWS comparison/analogue.

\revise{Additionally, there is a direct mapping of the phases of GDB operations to information exchanges in the MAC for device grouping, as noted on the left and bottom-right of Fig.~\ref{fig:gdb}. Specifically, the hierarchical aggregation of geolocation data from SLAVE devices and the distribution of GDB responses to them are carried out in the DA and DD slots, respectively. The direct communication between the GDB and the MASTER devices, in contrast, is executed over the BS-GC links during the AUT/ADT slots in uplink/downlink, respectively. The GDB may also exploit the guarding interval G to assess the situation and generate its response to sensors.} Here, the GDB signaling directly parallels concepts such as TVWS and CBRS, where the question of whether the exchange with the GDB is at the MAC as in this case, or at higher layers as would be conventionally the case in TVWS/CBRS, is somewhat academic. There is therefore a good argument for the broad GDB concept being extended to serve such purposes as grouping control in mMTC, as well as others that might be applicable in a 5G context such as rendezvous, spectrum/resource management and dynamic spectrum access, among others.


\section{Numerical Evaluation}\label{sec:simulation}
To verify our proposed method, we carry out an exhaustive numerical simulation campaign. The test scenario contains one base station serving a $200\times 200~\si{\meter}^2$ area. We consider a mixture of UEs with three different uplink transmission modes: $50\%$ aperiodic, $25\%$ with \SI{1}{\second} period and $25\%$ with \SI{10}{\second} period. All UEs are randomly assigned to $16$ different Access Classes (ACs) defined in LTE-A with even probabilities, where the UEs in ACs between $11$ and $15$ are considered as ultra-low-latency (ULL) devices. Additionally, we assume $50\%$ of the UEs to be static, and the remaining ones randomly walking at low mobility with a zero-mean speed and \SI{2}{\meter^2/\second^2} variance. UEs with higher levels of mobility are not taken into account as their presence are rare in the scenario of sensor networks. The assignments of uplink transmission mode, AC and mobility are independent from each other. We configure the RACH to a \SI{5}{\milli\second} slot length with $54$ available preambles per slot.

We consider a BLE implementation for our grouped RA approach with $20$ dBm transmitter antenna gain, $-90$ dBm as noise level, a maximal initial group size of $50$ UEs and an uplink package size of $64$ Bytes for periodic UEs. The clusters and group coordinators are initialized w.r.t. a-priori CSI by means to globally maximize the average SNR of D2D links, and then updated online according to the user feedback. As a benchmark, we also consider the EAB approach with an ACB barring factor of $0.1$, a maximal back-off time of \SI{0.5}{\second} and a System Information Block (SIB) broadcasting period of \SI{320}{\milli\second}. Both approaches are evaluated through $10$ times of Monte-Carlo test, each one simulating a \SI{30}{\second} period.
\begin{figure}[!htb]
	\centering
	\fbox{\includegraphics[width=.47\textwidth]{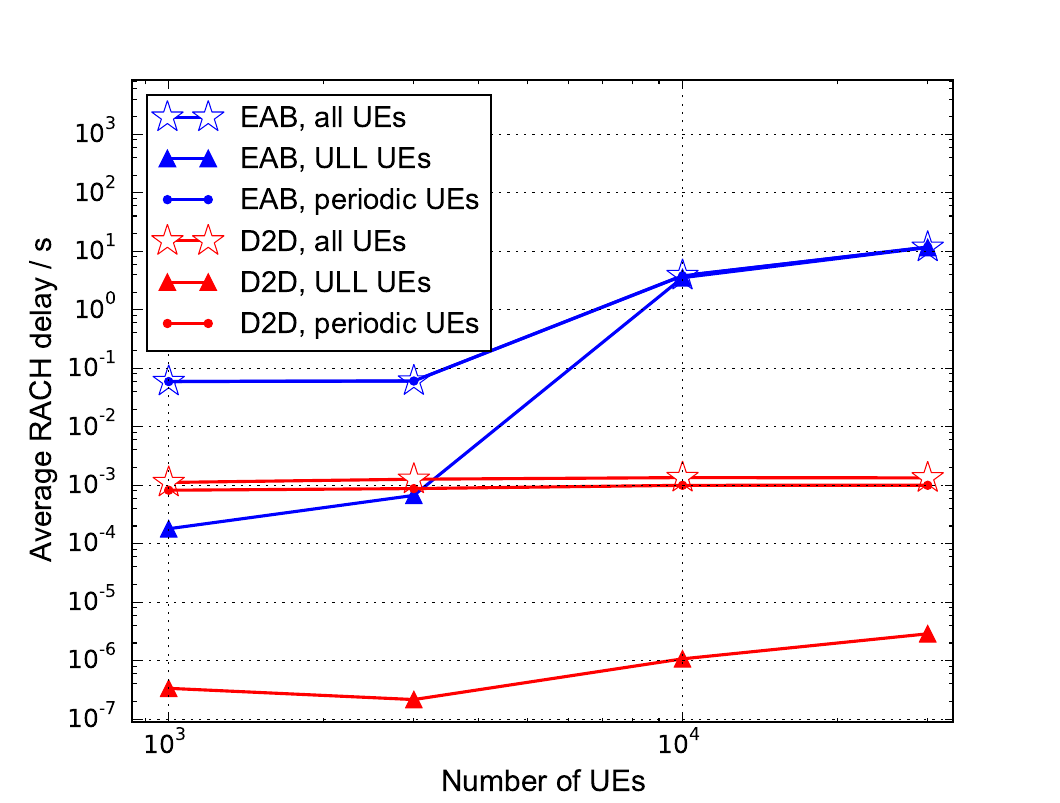}}
	\caption{Average RACH delay per RA attempt under EAB and our proposed approach}
	\label{fig:delay_comparison}
\end{figure}

The EAB approach fails to fulfill the ULL delay requirements when the UE number exceeds $10^4$, while our proposed method still remains effective even under a three-fold UE density, as shown in Fig.~\ref{fig:delay_comparison}. \revise{This empowers our solution and makes it ready for a real implementation in the upcoming next-generation networks not only tailored to sensor networks but also adapted to advanced 5G (and beyond) use cases.}

\section{Conclusion}\label{sec:conclusion}
The recently flourishing approaches of D2D-GRA are known to be attractive in sensor network environments with huge amount of devices, for their effective depression of data congestions and outstanding power efficiency. \revise{However, they have been generally developed only for environments with fully reliable D2D links that in some cases might not be a reasonable assumption.} To cope with this issue, in this paper we have thoroughly investigated their performance while deeply evaluating the D2D link reliability.
Through both analysis and simulations, we have shown that unreliable D2D links significantly hurt the performance of existing approaches in this category. Therefore, we have proposed an enhanced protocol that embeds extra signaling overhead into the user plane packets in order to reduce the increase in RA requests caused by D2D link exceptions and, in turn, to alleviate RAN data congestions. Finally, we have proposed a novel architectural concept: sensor grouping management through Geolocation Databases, yielding benefits in both cluster optimization and geolocation data collection.


%





\ifCLASSOPTIONcaptionsoff
  \newpage
\fi




\begin{thebibliography}{10}
	\providecommand{\url}[1]{#1}
	\csname url@samestyle\endcsname
	\providecommand{\newblock}{\relax}
	\providecommand{\bibinfo}[2]{#2}
	\providecommand{\BIBentrySTDinterwordspacing}{\spaceskip=0pt\relax}
	\providecommand{\BIBentryALTinterwordstretchfactor}{4}
	\providecommand{\BIBentryALTinterwordspacing}{\spaceskip=\fontdimen2\font plus
		\BIBentryALTinterwordstretchfactor\fontdimen3\font minus
		\fontdimen4\font\relax}
	\providecommand{\BIBforeignlanguage}[2]{{%
			\expandafter\ifx\csname l@#1\endcsname\relax
			\typeout{** WARNING: IEEEtran.bst: No hyphenation pattern has been}%
			\typeout{** loaded for the language `#1'. Using the pattern for}%
			\typeout{** the default language instead.}%
			\else
			\language=\csname l@#1\endcsname
			\fi
			#2}}
	\providecommand{\BIBdecl}{\relax}
	\BIBdecl
	

	\bibitem{soldani2015horizon}
	D.~Soldani and A.~Manzalini, ``Horizon 2020 and beyond: On the 5G operating system for a true digital society,'' \emph{IEEE Vehicular Technology Magazine}, 10.1 (2015): 32--42.
	
	\bibitem{3gpp2014geran}
	``3GPP TR 43.868 V12.0.0, GERAN Improvements for Machine-Type Communications,'' Technical Report, 3GPP, Feb. 2014.
	
	\bibitem{hasan2013random}
	M.~Hasan, E.~Hossain, and D.~Niyato, ``Random access for Machine-to-Machine Communication in LTE-Advanced networks: Issues and approaches,'' \emph{IEEE	Communications Magazine}, 51.6 (2018): 86--93.
	
	\bibitem{ali2017lte}
	M.~S. Ali, E.~Hossain and D.~I. Kim, ``LTE/LTE-A random access for massive Machine-Type Communications in smart cities,'' \emph{IEEE Communications	Magazine}, 55.1 (2017): 76--83.

	\bibitem{lien20163gpp}
	S.~Y.~Lien, C.~C.~Chien, F.~M.~Tseng, and T.~C.~Ho, ``3GPP Device-to-Device Communications for beyond 4G cellular networks'', \emph{IEEE Communications Magazine}, 54.3 (2016): 29--35.

	\bibitem{wu2016device}
	Y.~Wu, W.~Guo, H.~Yuan, et al., ``Device-to-Device meets LTE-Unlicensed'', \emph{IEEE Communications Magazine}, 54.5 (2016): 154--159.

	
	\bibitem{sciancalepore2016offloading}
	V.~Sciancalepore, D.~Giustiniano, A.~Banchs, and A.~Hossmann-Picu, ``Offloading cellular traffic through opportunistic communications: Analysis and optimization,'' \emph{IEEE Journal on Selected Areas in Communications}, 34.1 (2016): 122--137.

	\bibitem{soldani2008wireless}
	\revise{D.~Soldani, and S.~Dixit, ``Wireless relays for broadband access [radio communications series],'' \emph{IEEE Communications Magazine} 46.3 (2008): 58--66.}

	\bibitem{ji2017applying}
	L.~ Ji, B.~Han, M.~Liu and H.~D.~Schotten, ``Applying Device-to-Device communication to enhance IoT services'', \emph{IEEE Communications Standards Magazine}, 1.2 (2017): 85--91.
	
	\bibitem{chatzikokolakis2015way}
	K.~Chatzikokolakis, A.~Kaloxylos, P.~Spapis, et al., ``On the way to massive access in 5G: Challenges and solutions for massive machine communications,'' in \emph{International Conference on Cognitive Radio Oriented Wireless Networks}, 2015: 708--717.




	
	\bibitem{3gpp2015study}
	``3GPP TS 38.913 V14.3.0, Study on Scenarios and Requirements for Next Generation Access Technologies,'' Technical Specification, 3GPP, Dec. 2015.
	
	\bibitem{han2017group}
	B.~Han and H.~D. Schotten, ``Grouping-based random access collision control for massive Machine-Type Communication,'' \emph{2017 IEEE Global Communications Conference}, Singapore, 2017.
	
	\bibitem{hol2016}
	O.~Holland, ``Some are born with white space, some achieve white space, and some have white space thrust upon them,'' \emph{IEEE Transactions on Cognitive Communications and Networking}, 2.2 (2016): 178--193.
	
	\bibitem{mat2014}
	M.~Matinmikko, H.~Okkonen, M.~Palola, et al., ``Spectrum sharing using licensed shared access: The concept and its workflow for LTE-Advanced networks,'' \emph{IEEE Wireless Communications}, 21.2 (2014): 72--79.
	
	\bibitem{CBRS2015}	``In the Matter of Amendment of the Commission's Rules with Regard to Commercial Operations in the 3550-3650 MHz Band, Report and Order and Further Notice of Second Proposed Rulemaking,'' Technical Document, FCC, Apr. 2015.
	
\end{thebibliography}
%

%
%
\begin{IEEEbiography}[{\includegraphics[width=1in,height=1.25in,clip,keepaspectratio]{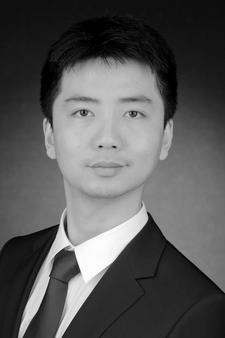}}]{
Bin Han} (M'15) received in 2009 his B.E. degree from Shanghai Jiao Tong University, M.Sc. in 2012 from Technische Universit\"at Darmstadt, and in 2016 the Dr.-Ing. degree from Kalsruhe Institute of Technology. Since July 2016 he has been with Technische Universit\"at Kaiserslautern, researching in the broad area of wireless networks and signal processing.
\end{IEEEbiography}


\begin{IEEEbiography}[{\includegraphics[width=1.0in,height=1.25in,clip,keepaspectratio]{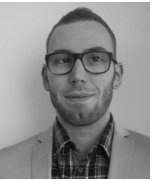}}]{
Vincenzo Sciancalepore} (S'11--M'15) received his M.Sc. degree in Telecommunications Engineering and Telematics Engineering in 2011 and 2012, respectively, whereas in 2015, he received a double Ph.D. degree. Currently, he is a senior 5G researcher at NEC Laboratories Europe, focusing on network virtualization and network slicing challenges. 
\end{IEEEbiography}


\begin{IEEEbiography}[{\includegraphics[width=1.0in,height=1.25in,clip,keepaspectratio]{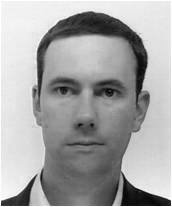}}]{
		Oliver Holland} (S'02--M'06) is a senior researcher at King's College London. He is an internationally-renowned expert on TVWS and other spectrum sharing technologies, spectrum in general, and upcoming 5G wireless communication technologies, among other topics.
\end{IEEEbiography}


\begin{IEEEbiography}[{\includegraphics[width=1.0in,height=1.25in,clip,keepaspectratio]{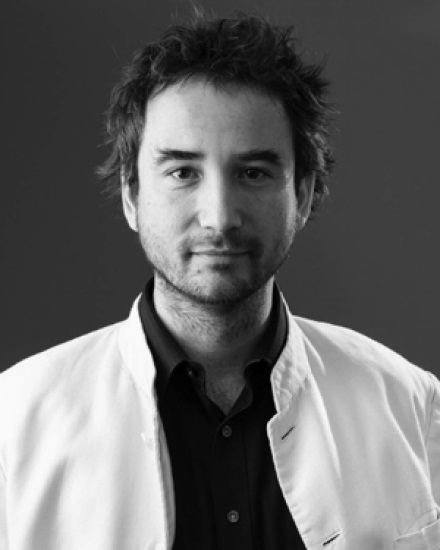}}]{
Mischa Dohler} (S'99--M'03--SM'07--F'14) is currently Full Professor in wireless communications with King's College London, the Head of the Centre for Telecommunications Research. He was a Distinguished Lecturer of the IEEE. He was the Editor-in-Chief of the Transactions on Emerging Telecommunications Technologies and the Transactions on the Internet of Things.
\end{IEEEbiography}


\begin{IEEEbiography}[{\includegraphics[width=1in,height=1.25in,clip,keepaspectratio]{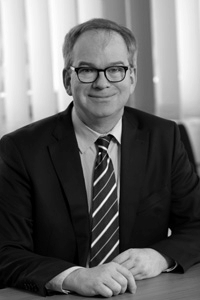}}]{
Hans D. Schotten} (S'93--M'97) received the Diplom and Ph.D. degrees in Electrical Engineering from RWTH Aachen in 1990 and 1997, respectively. Since 2007, he has been Full Professor and Head of the Institute of Wireless Communication at the Technische Universit\"at Kaiserslautern. Since 2012, he has been Scientific Director at the German Research Center for Artificial Intelligence heading the ''Intelligent Networks'' department.
\end{IEEEbiography}

\end{document}